\documentclass[journal,transmag]{IEEEtran}
\usepackage{hyperref,color,breqn,multirow}
\usepackage[top=0.7in, left=0.65in]{geometry}
\usepackage{graphicx}
\usepackage{siunitx}
\usepackage{tudacolors}

\renewcommand{\vec}[1]{\mathbf{#1}}
\newcommand{\onref}[1]{\hat{#1}}

\usepackage{eso-pic}
\AddToShipoutPicture*{\footnotesize\sffamily\raisebox{1cm}{\hspace{1.65cm}\fbox{\parbox{\textwidth}{\copyright 2022 IEEE.  Personal use of this material is permitted.  Permission from IEEE must be obtained for all other uses, in any current or future media, including reprinting/republishing this material for advertising or promotional purposes, creating new collective works, for resale or redistribution to servers or lists, or reuse of any copyrighted component of this work in other works.}}}}

\begin{document}
\title{Torque Computation with the Isogeometric Mortar Method\\for the Simulation of Electric Machines}
\author{\IEEEauthorblockN{Melina Merkel\IEEEauthorrefmark{1,2},
Bernard Kapidani\IEEEauthorrefmark{3}, Sebastian Schöps\IEEEauthorrefmark{1,2}, and
Rafael Vázquez\IEEEauthorrefmark{3}}\IEEEauthorblockA{\IEEEauthorrefmark{1}Computational Electromagnetics Group, Technische Universität Darmstadt, 64289 Darmstadt, Germany}
\IEEEauthorblockA{\IEEEauthorrefmark{2}Centre for Computational Engineering, Technische Universität Darmstadt, 64289 Darmstadt, Germany}
\IEEEauthorblockA{\IEEEauthorrefmark{3}Chair of Numerical Modelling and Simulation, École Polytechnique Fédérale de Lausanne, 1015 Lausanne, Switzerland}
}

\IEEEtitleabstractindextext{\begin{abstract}
In this work isogeometric mortaring is used for the simulation of a six pole permanent magnet synchronous machine. Isogeometric mortaring is especially well suited for the efficient computation of rotating electric machines as it allows for an exact geometry representation for arbitrary rotation angles without the need of remeshing. The appropriate B-spline spaces needed for the solution of Maxwell's equations and the corresponding mortar spaces are introduced. Unlike in classical finite element methods their construction is straightforward in the isogeometric case. The torque in the machine is computed using two different methods, i.e., Arkkio's method and by using the Lagrange multipliers from the mortaring.
\end{abstract}

}

\maketitle
\thispagestyle{empty}
\pagestyle{empty}

\section{Introduction}
Due to the rise of e-mobility and the increased use of electric motors, there is large need for the simulation of machines. A bottleneck in the design workflow is the treatment of geometry and meshing.  Up to \SI{75}{\percent} of the simulation time is spent on modeling and geometry discretization \cite{Boggs_2005aa}. The reason is that most simulations are carried out using the classical Finite Element Method (FEM) which uses element-wise polynomial mappings of (curved) triangulations of the geometry. Those mappings cannot exactly describe conic sections and since common machines have a cylindrical structure, a geometry error is introduced. This is particularly inconvenient in the air-gap where the fields are of highest interest. 

Commonly, rotation is enabled by a cut in the air gap and a domain decomposition into rotor and stator is applied. Many methods have been proposed in the literature for handling the resulting non-matching discretizations, e.g.~`locked step', `sliding surface', `moving band' and `harmonic coupling' approaches, see for example \cite{De-Gersem_2004ad,Davat_1985aa,Lai_1992aa,Wallinger_2019aa}. Also, boundary element coupling is a viable option to consider motion \cite{Kurz_1995aa}. Finally, Nitsche \cite{Roppert_2020aa} and mortaring methods \cite{Buffa_1999aa,Bouillault_2001aa} have been proposed. In particular, mortaring is commonly considered to be more complex in its implementation and computationally heavier at run-time, but accounts for the trace spaces in a mathematically correct way.

\begin{figure}
	\centering
	\scalebox{0.9}{
	\includegraphics{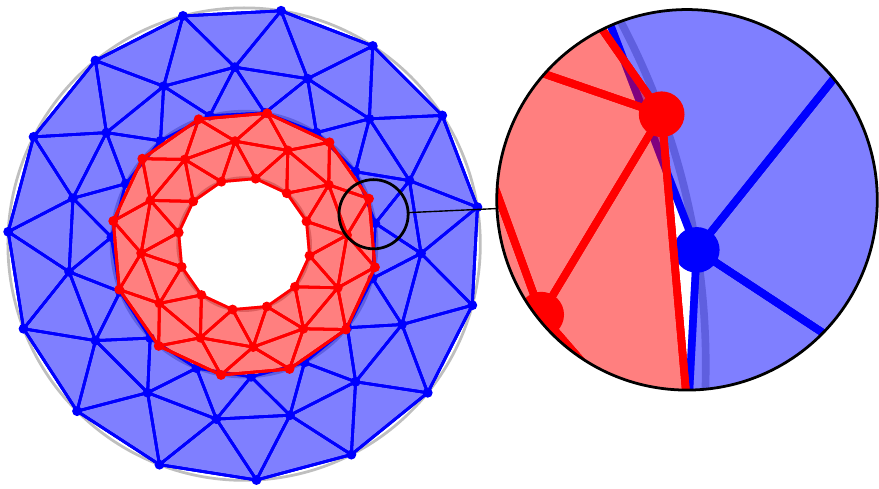}}
	\caption{Schematic visualization of element mismatch at the interface between rotor (red) and stator (blue)  the circular interface is not exactly represented by the mesh.}
	\label{fig:rotoriga}
\end{figure}

All `sliding' methods share the the problem of geometric mismatch at the interface, see \autoref{fig:rotoriga}. This may lead to numerical noise which is particular cumbersome when analyzing cogging torque. When using conventional finite elements this can be mitigated by mesh adaptation, see \cite[Section 5.4]{Lange_2012ab} and similarly \cite[Section 5.2.3]{Wallinger_2019aa}. The geometry error can be avoided when using Isogeometric Analysis (IGA) \cite{Hughes_2005aa} which uses B-Splines and Non-Uniform Rational B-Splines (NURBS) for both the geometry representation (as weighting functions on a structured list of control points, defining a control mesh) and as basis functions for test-spaces and discretization. Classically, a conforming multipatch discretization is needed for IGA \cite{Buffa_2015aa}. While IGA has been used for machine models in the past, e.g., \cite{Bontinck_2018ac,Friedrich_2020aa,Merkel_2021ab}, this paper presents the first isogeometric simulation workflow for rotating machines with focus on torque computation. To implement rotation efficiently, we extend the isogeometric mortar method from \cite{Buffa_2020aa} and use a tree-cotree regularization~\cite{Kapidani_2021aa}.

The paper is structured as follows; we discuss the mathematical model in Section~\ref{sec:math} and the discretization using IGA in Section~\ref{sec:iga}.
Mortaring and tree-cotree regularization are introduced in Sections~\ref{sec:mortar} and \ref{sec:treecotree}, respectively. Different methods for torque computation are given in Section~\ref{sec:torque}. Finally, numerical results are presented in Section~\ref{sec:numerics} and the paper is closed by conclusions in Section~\ref{sec:conclusion}.

\section{Mathematical Machine Model}\label{sec:math}
The electromagnetic behavior of electric machines can be described by Maxwell's equations. Under common assumptions, the magnetostatic approximation can be employed. Using a magnetic vector potential $\mathbf{A}$, such that $\mathbf{B}=\operatorname{curl}\mathbf{A}$, the problem in a 3D domain $\Omega$ can be written as
\begin{align}
\operatorname{curl}\left(\nu \operatorname{curl} \mathbf{A} \right) = \mathbf{J}_{\mathrm{src}} + \operatorname{curl} \mathbf{M} , \label{eq:curlcurl} \end{align}
where $\nu$ is the magnetic reluctivity, $\mathbf{J}_{\mathrm{src}}$ is a source current density and $\mathbf{M}$ is the magnetization of permanent magnets.
The weak formulation can be derived as: find $\mathbf{A}\in V$, s.t.
\begin{equation}\label{eq:weakMagnetostatic}
    \begin{aligned}
        (\nu  \operatorname{curl} \mathbf{A}  ,  \operatorname{curl} \mathbf{v} )_{\Omega} - ( \nu \operatorname{curl} \mathbf{A} \times \mathbf{n}  , \mathbf{v} )_{\partial\Omega}  \\
        =
        ( \mathbf{J}_{\mathrm{src}},\mathbf{v})_{\Omega} - (\mathbf{M}  , \operatorname{curl} \mathbf{v} )_{\Omega},
    \end{aligned}
\end{equation}
for all basis functions $\mathbf{v} \in V$, where $V \subset {H}(\mathrm{curl};\Omega)$, fulfill the boundary conditions on $\partial\Omega$. 

Let us consider the case of a circular rotating machine. We introduce a cut within the air gap $\Omega_{\mathrm{air}}$ decomposing the domain into $\Omega=\Omega_1\cup\Omega_2$ for stator/rotor. We call the interface $\Gamma_{12}=\partial\Omega_1 \cap \partial\Omega_2 \subset \Omega_{\mathrm{air}}$ with normal vector $\mathbf{n}_{12}$ and denote the air gap pieces by $\Omega_{\mathrm{air},i} = \Omega_i\cap\Omega_{\mathrm{air}}$. The angle $\alpha$ encodes the angular offset between rotor and stator due to motion. Then, we may choose any closed surface $S$ that is contained in the air gap, e.g., the lateral cylinder surface $S=\Gamma_{12}$ to compute the torque, \cite[p. 55f]{Arkkio_1987aa}. It can be parametrized conveniently by a cylindrical coordinate system with (constant) radial distance $r=R$, angle $\theta\in[0,2\pi]$ and axial coordinate $z\in[0,l_z]$ where $l_z$ is the length of the machine. Now, the torque can be computed as the surface integral, see e.g.~\cite[eq. (121)]{Arkkio_1987aa} 
\begin{equation}\label{eq:torque}
T_\theta(\alpha) = \int\limits_0^{l_z}\int\limits_0^{2\pi} B_r(\alpha) H_\theta(\alpha) R^2\; \mathrm{d}\theta\;\mathrm{d}z,
\end{equation}
where $B_r(\alpha)$ and $H_\theta(\alpha)$ are the radial and angular components of the magnetic flux density $\mathbf{B}$ and field strength $\mathbf{H}=\nu\mathbf{B}$ for each rotor angle $\alpha$, respectively. Note, that we dropped the spatial coordinates and transforms for readability. 
\begin{figure}
	\centering
	\includegraphics{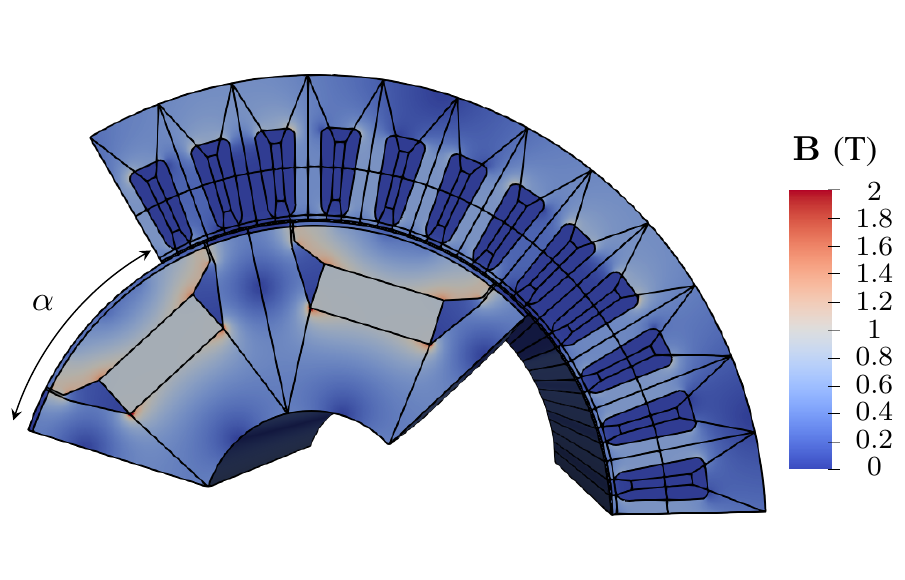}
	\caption{Magnetic flux density in two poles of a six-pole permanent magnet synchronous machine with angle $\alpha$, computed using isogeometric mortaring.}
	\label{fig:machine}
\end{figure}

\section{Isogeometric Analysis}\label{sec:iga}
The geometry is defined by one or several trivariate \emph{B-splines} or NURBS mappings \cite{Cohen_2001aa} from a reference domain $\hat{\Omega}=(0,1)^{3}$ to the physical domain $\Omega$. For geometries which cannot be represented by a regular mapping a conforming multipatch model can be used where the geometry of each patch is defined by a separate trivariate mapping. 
Following the usual notation, B-splines of degree $p$ are defined from the knot vector
\begin{equation}
	\label{eq:knot}
	\Xi = \left[\xi_1, \dots, \xi_{n+p+1}\right] \quad 0 \leq \xi_i \leq \xi_{i+1} \leq 1,
\end{equation}
using the Cox-De Boor recursion formula~\cite{Cohen_2001aa}, where $n$ is the number of functions. Let us denote by $\onref{B}_i^p$ the $i$-th basis function of degree $p$ on the reference domain $(0,1)$ and define the trivariate space of B-splines as
\begin{equation}
\begin{aligned}
S_{p_1,p_2,p_3}&(\Xi_1,\Xi_2,\Xi_3)
	=\\
	&\mathrm{span} 
	\left\lbrace
	\prod_{k=1}^3\onref{B}_{i_k}^{p_k}(\Xi_k)
\;\Big|\;
	1 \leq i_k \leq n_k
	\right\rbrace.
\end{aligned}
\end{equation}

As we need edge element basis functions for the solution of \eqref{eq:weakMagnetostatic}, we require a curl-conforming spline space.
For simplicity we assume that the polynomial degree is the same in all parametric directions, i.e., $p = p_1 = p_2 = p_3$. Following~\cite{Buffa_2020aa}, we define the corresponding spline space in the reference domain~$\onref{\Omega}$ as
\begin{equation}
\begin{aligned}
S_p^1(\onref{\Omega})&= S_{p-1,p,p}(\Xi_1',\Xi_2,\Xi_3)\\
	&\times S_{p,p-1,p}(\Xi_1,\Xi_2',\Xi_3)\\
	&\times 	S_{p,p,p-1}(\Xi_1,\Xi_2,\Xi_3'),
	\label{eq:S1hat}
\end{aligned}
\end{equation}
where $\Xi_j'=[\xi_2^j,\dots,\xi_{n_j+p}^j]$ for $j=1,2,3$ is a modified knot vector with the first and last knot removed. This corresponds to lowering the degree and regularity of the basis functions by one in the $j$-th direction. 
Then, considering the geometry mappings and their Jacobians, the multipatch spline spaces in the physical domain $\Omega$ are defined by push-forwards. We obtain
\begin{equation}
	V := S_p^1(\Omega),
	\quad
	M_{12} := \boldsymbol{\gamma}_{\perp,\Gamma_{12}}(S_q^1(\Omega_1)), \end{equation}
with the trace operator $\boldsymbol{\gamma}_{\perp,\Gamma_{12}}(\mathbf{u}) := \mathbf{u} \times \mathbf{n}_{12}$ restricted to $\Gamma_{12}$. A stable mortar space $M_{12}$ for the multiplier can be easily created by lowering its degree by an odd number with respect to the edge-element space of the vector potential, e.g., $q=p-1$ by removing knots from \eqref{eq:knot}. Boundary conditions on $\partial\Omega$ may still be implemented. Note that these spline spaces are part of a de Rham complex \cite{Buffa_2020aa}.
\subsection{Mortaring}\label{sec:mortar}
Stator $\Omega_{1}$ and rotor $\Omega_{2}$ are modeled independently and are glued for angle $\alpha$ using our isogeometric mortar method \cite{Buffa_2020aa}. Note that we will drop the $\alpha$ dependency for readability if possible.
The coupling is realized by introducing a Lagrange multiplier
\begin{align*}
    \boldsymbol{\lambda} = \nu \operatorname{curl} \mathbf{A} \times \mathbf{n}_{12},
\end{align*}
to \eqref{eq:weakMagnetostatic}, where $\mathbf{n}_{12}$ is the normal vector on $\Gamma_{12}$, yielding the following problem: find $\mathbf{u} \in V$, $\boldsymbol{\boldsymbol{\lambda}} \in M_{12}$ such that
\begin{align} 
	\label{eq:weak1}
	a\bigl(\mathbf{u}, \mathbf{v}\bigr)_{\Omega}
	+
	b\bigl(\mathbf{v}, \boldsymbol{\lambda}\bigr)_{\Gamma_{12}} 
	&=
	{f}(\mathbf{v})_{\Omega}
	\\
	\label{eq:weak2}
	b\bigl(\mathbf{u}, \boldsymbol{\mu}\bigr)_{\Gamma_{12}} 
	&=
	0
\end{align}
	for all $\mathbf{v} \in V$ and $\boldsymbol{\mu} \in M_{12}$ and the bilinear forms
\begin{align}
	\label{eq:curlcurl_bf}
	a(\mathbf{u}, \mathbf{v})_{\Omega} &:= \sum_{k=1}^{2} (\nu\operatorname{curl} \mathbf{u}, \operatorname{curl} \mathbf{v})_{\Omega_{k}},
	\\
	\label{eq:quadrature}
	b(\mathbf{u}, \boldsymbol{\mu})_{\Gamma_{12}} &:=  ([\boldsymbol{\gamma}_{12}(\mathbf{u})], \boldsymbol{\mu})_{\Gamma_{12}} \intertext{and right-hand-side}
	{f}(\mathbf{v})_{\Omega}
	&:=
	\sum_{k=1}^{2} 
		\bigl(\mathbf{J}_{\mathrm{src}},\mathbf{v}\bigr)_{\Omega_{k}} - \bigl(\mathbf{M} , \operatorname{curl}\mathbf{v}\bigr)_{\Omega_{k}}
\end{align}
where $\boldsymbol{\gamma}_{12}(\vec{u})=(\mathbf{n}_{12}\times\vec{u})\times\mathbf{n}_{12}$ and the brackets $[\cdot]$ denote the jump across the interface $\Gamma_{12}$. 

\begin{figure}
\centering
\includegraphics{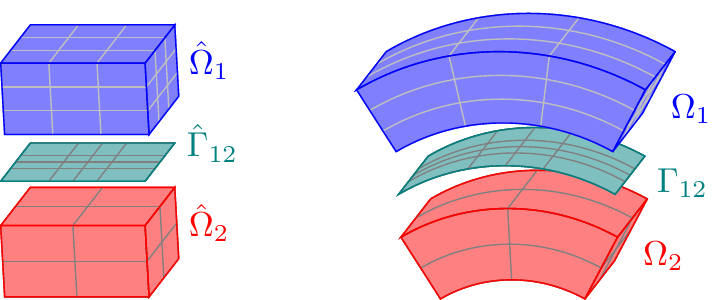}
\caption{Schematic visualization of the mesh of the reference domains $\onref{\Omega}_{1}$, $\onref{\Omega}_{2}$ and the intersection mesh on the interface $\onref{\Gamma}_{12}$ on the left and of the physical domains ${\Omega}_{1}$, ${\Omega}_{2}$ and the intersection mesh on the interface ${\Gamma}_{12}$ on the right.} \label{fig:intersectionmesh}
\end{figure}

Spectral correctness and inf-sup stability of the saddle-point problem (\ref{eq:weak1}-\ref{eq:weak2}) for adequate choices of the spaces have been recently proven in \cite{Buffa_2020aa}.
This is a significant advantage of isogeometric mortaring compared to other mortar variants, e.g., using conventional finite elements with rather complicated constructions of dual basis functions. One key issue is the construction of an intersection mesh in the reference domain $\onref{\Gamma}$ for the numerical quadrature of the integral \eqref{eq:quadrature}, see Fig.~\ref{fig:intersectionmesh}. This is also straightforward due to the tensor product structure of the spaces.

\subsection{Tree-Cotree Regularization}\label{sec:treecotree}
The removal of the (discrete) kernel of the curl operator is a necessary requirement for the well-posedness of formulation \eqref{eq:weak1}--\eqref{eq:weak2}. In particular, the matrix discretizing \eqref{eq:curlcurl_bf} will be singular on each subdomain of the domain decomposition.
A popular way to amend the issue in low order FEM ($p=1$) is to set the solution a priori on the spanning tree of edges in the vertex-edge mesh graph. This works because of the discrete degrees of freedom (DoFs) correspondence to line integrals on edges of the FEM mesh, which in turn easily permits to identify the gradient subspace of $V$.

Unfortunately, in the IGA setting just as in FEM, as soon as $p>1$, the DoFs of $V$ will lose their physical interpretation. Nevertheless, as noted in the literature~\cite{Beirao-da-Veiga_2014aa} the basis functions of the spline spaces $S_p^k(\Omega)$ in the discrete de Rham sequence can be identified with $k-$dimensional entities of the underlying piecewise affine control mesh, i.e., with vertices, edges, facets, and cells for $k=0,1,2$ and $3$, respectively.
Because of these one-to-one correspondences, it suffices to apply the algorithm for the tree and cotree construction to the control mesh graph itself, and identify the corresponding B-spline functions. Very conveniently, this observation holds also for different boundary conditions and in the case of non-contractible domains. In this last case we can apply the techniques used to identify generators of cohomology groups for low order schemes in \cite{Dlotko_2017aa} to the standard tree-cotree decomposition. Finally, the extension to mortaring is described in \cite{Kapidani_2021aa}.

\section{Torque Calculation}\label{sec:torque}
In our mortaring setting, one easily verifies that the third component of the Lagrange multiplier $\lambda_3=-H_\theta$ can be identified with the (negative) angular component of the magnetic field strength. Therefore, we can immediately compute the torque by
\begin{equation}\label{eq:torque_mortar}
	T_\theta(\alpha) = -\int\limits_0^{l_z}\int\limits_0^{2\pi} B_r(\alpha) \lambda_3(\alpha) R^2\; \mathrm{d}\theta\;\mathrm{d}z
\end{equation}
when evaluating $B_r=\operatorname{curl}(\mathbf{A})\cdot\mathbf{n}_{12}$ either on the boundary of $\Omega_1$ (stator) or $\Omega_2$ (rotor). Note that the corresponding numerical quadrature (on patch boundaries in the reference domain $\onref{\Omega}$) is straight forward since we are integrating on a perfect cylindrical shape. However, such direct evaluation of \eqref{eq:torque} is commonly considered numerically unstable \cite[p. 55]{Arkkio_1987aa}, i.e., the result is very sensitive with respect to the choice of the integration interface $S$. Therefore a large variety of methods and formulae to calculate the electromagnetic torque has been developed, a comprehensive literature review is given in \cite{Henrotte_2010aa}. A particular popular variant is Arkkio's method which averages the torque over some volume, i.e.,
\begin{equation}\label{eq:torque_arkkio}
T_\theta(\alpha) = \int\limits_0^{l_z} \int\limits_{0}^{\delta}\int\limits_0^{2\pi} B_r(\alpha) H_\theta(\alpha) (R+r)^2\; \mathrm{d}\theta\;\mathrm{d}r\;\mathrm{d}z.
\end{equation}
where we can, for example, evaluate the fields either in the air gap contained in $\Omega_{\mathrm{air},1}$ (stator) or $\Omega_{\mathrm{air},2}$ (rotor).

\begin{figure}
	\includegraphics{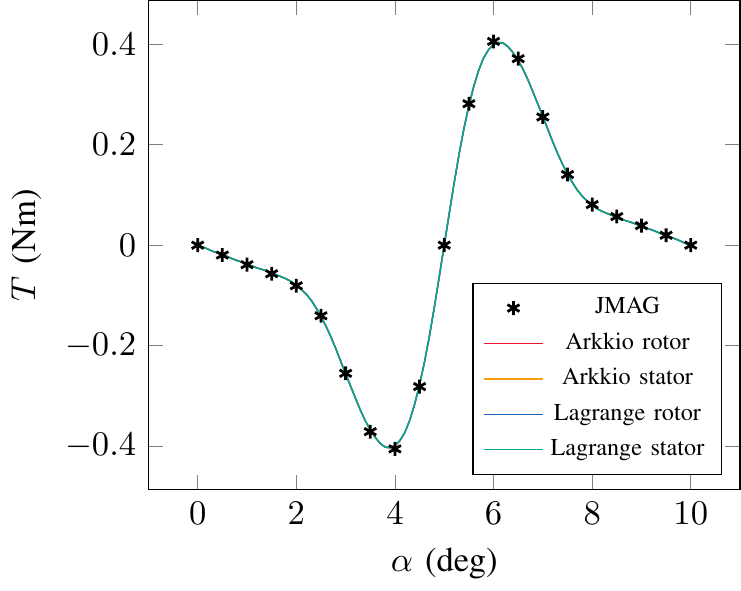}
	\vspace{-0.5em}
	\caption{Torque computations via
		reference 2D FE solution (JMAG),
		Arkkio's method (\ref{eq:torque_arkkio}) on rotor and stator side (Arkkio rotor/stator),
		Lagrange multipliers method (\ref{eq:torque_mortar}) on rotor and stator side (Lagrange rotor/stator).
	}\label{fig:torque}
\end{figure}

\section{Numerical Results}\label{sec:numerics}
\begin{figure}
	\includegraphics{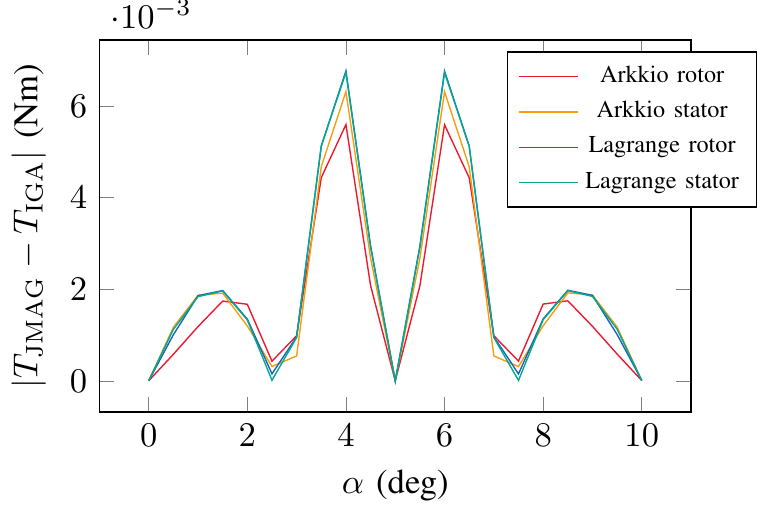}
	\vspace{-0.5em}
	\caption{Absolute difference between the torque computed via
		reference 2D FE solution (JMAG) and the torque computed with IGA using
		Arkkio's method (\ref{eq:torque_arkkio}) on rotor and stator side (Arkkio rotor/stator),
		Lagrange multipliers method (\ref{eq:torque_mortar}) on rotor and stator side (Lagrange rotor/stator).
	}\label{fig:torque_diff}
\end{figure}

The isogeometric mortaring described in (\ref{eq:weak1}-\ref{eq:weak2}) is used to compute the magnetic field of a six-pole permanent magnet synchronous machine, whose details are given in \cite{Merkel_2021ab}. Degree $p=2$ is used for the discrete edge element spaces $V$ and $q=1$ for the multiplier space $M_{12}$ which is defined on the interface mesh of $\Omega_1$. The rotor consists of 24 patches, resulting in 540353 cotree DoFs (and 309663 eliminated tree DoFs), the stator is given by 228 patches and 1058880 cotree DoFs (and 601392 eliminated tree DoFs), the multiplier adds another 3912 DoFs. The model and magnetic flux density in this machine can be seen in Fig.~\ref{fig:machine}. The cogging torque is computed using the Lagrange multipliers (\ref{eq:torque_mortar}) and Arkkio's method (\ref{eq:torque_arkkio}) on $\Omega_1$ (stator) and $\Omega_2$ (rotor), respectively. In Arkkio's method the integration volume is chosen in both cases as the whole air gap piece $\Omega_{\mathrm{air},i}$. The results are shown in Fig.~\ref{fig:torque} and compared to a 2D reference solution computed using JMAG\textsuperscript{\textregistered} with a finite element discretization of order $p=1$ with 1618173 elements. The difference between the torque computed with the IGA approaches and the JMAG reference solution is shown in Fig.~\ref{fig:torque_diff}. All approaches are in excellent agreement. We do not observe the often reported instability for surface/line evaluation of the torque. In our approach, the cylindrical shape of the machine can be exactly described and the angular component of the magnetic field can be obtained from the Lagrange multipliers alleviating the need to numerically differentiate the magnetic vector potential for this component.  

\section{Conclusion}\label{sec:conclusion}
We have introduced isogeometric mortaring for rotating electric machines using a tree-cotree decomposition for the regularization of the system. Two different methods for the torque computation of electric machines have been introduced and used for the computation of the cogging torque of a 3D six-pole permanent magnet synchronous machine. Isogeometric analysis and in particular isogeometric mortaring were demonstrated to be very well suited for the torque computation. Surface and volume integration for the evaluation of the torque are in excellent agreement with a reference solution.

\section*{Acknowledgements}
This work is supported by the Graduate School CE within the Centre for Computational Engineering at Technische Universität Darmstadt, by the Swiss National Science Foundation via the project HOGAEMS n.200021\_188589, and by the Defense Advanced Research Projects Agency (DARPA), under contract HR0011-17-2-0028. The views, opinions and/or findings expressed are those of the author and should not be interpreted as representing the official views or policies of the Department of Defense or the U.S. Government. We thank the JSOL Corporation for the fruitful discussions, partnership and JMAG license.

\end{document}